\documentclass[preprint]{vgtc}               

\ifpdf
  \pdfoutput=1\relax                   
  \usepackage{graphicx}                
  \DeclareGraphicsExtensions{.pdf,.png,.jpg,.jpeg} 
\else
  \usepackage{graphicx}                
  \DeclareGraphicsExtensions{.eps}     
\fi%

\graphicspath{{figs/}{pictures/}{images/}{./}} 

\usepackage{microtype}                 
\PassOptionsToPackage{warn}{textcomp}  
\usepackage{textcomp}                  
\usepackage{mathptmx}                  
\usepackage{times}                     
\usepackage{cite}                      
\usepackage{tabu}                      
\usepackage{booktabs}                  

\usepackage{soul}
\usepackage{algorithm}
\usepackage{algpseudocode}
\usepackage{enumitem}
\usepackage{amsmath}
\usepackage{url}
\usepackage{subfig}
\usepackage[strings]{underscore}
\usepackage{color}
\usepackage{subfig}
\usepackage[font=sf]{caption}

\newenvironment{tightItemize}{
\begin{itemize}[topsep=3pt]
        \setlength{\itemsep}{1pt}
        \setlength{\parskip}{0pt}
        \setlength{\parsep}{0pt}
}{\end{itemize}
}

\newenvironment{tightEnumerate}{
\begin{enumerate}[topsep=3pt]
        \setlength{\itemsep}{1pt}
        \setlength{\parskip}{0pt}
        \setlength{\parsep}{0pt}
}{\end{enumerate}
}

\onlineid{1018}

\vgtccategory{LDAV 2018 Papers}

\vgtcinsertpkg

\title{DPP-PMRF: Rethinking Optimization for a Probabilistic Graphical Model Using Data-Parallel Primitives}

\author{Brenton Lessley\thanks{e-mail: blessley@cs.uoregon.edu}\\ %
        \scriptsize University of Oregon %
\and Talita Perciano
    \\ %
     \scriptsize Lawrence Berkeley Nat'l Lab %
\and Colleen Heinemann
    \\ %
	 \scriptsize Lawrence Berkeley Nat'l Lab %
\and David Camp
    \\ %
	 \scriptsize Lawrence Berkeley Nat'l Lab %
\and Hank Childs
    \\ %
	 \scriptsize University of Oregon %
\and E. Wes Bethel
    \\ %
	 \scriptsize Lawrence Berkeley Nat'l Lab %
     }

\abstract{
We present a new parallel algorithm for probabilistic graphical model optimization.
The algorithm relies on data-parallel primitives (DPPs), which provide portable performance over hardware architecture.
We evaluate results on CPUs and GPUs for an image segmentation problem.
Compared to a serial baseline, we observe runtime speedups of up to 13X (CPU) and 44X (GPU).
We also compare our performance to a reference, OpenMP-based algorithm, and find speedups of up to 7X (CPU).

} 

\CCScatlist{
  \CCScatTwelve{Mathematics of computing}{Graph algorithms, Markov networks, Expectation maximization}{}{};
  \CCScatTwelve{Computing methodologies}{Image processing, Shared memory}{}{}
}

\begin{document}



\maketitle

\section{Introduction}


Image segmentation refers to labeling regions of an image based on specific pixel properties.
There are many approaches for this task, resulting in different types of segmentations.
One approach is to transform an image into a graph, where each region of pixels is encoded as a vertex.
For each vertex (i.e., region of pixels), there are multiple possible labels, and each label is assigned a probability.
The labels are chosen by minimizing an energy function, which is solved via an optimization routine.
This entire process is referred to as \textit{probabilistic graphical model} (PGM) optimization.
In this study, we consider a specific form of this process---described in detail in Section \S\ref{sec:background}---that uses \textit{Markov Random Fields} (MRFs).

This process, i.e., constructing the graph and optimizing the energy function, requires significant computation.
For our use cases and data sets, serial execution times were not sufficient.
%
This work focuses on a novel reformulation of MRF optimization 
for shared-memory parallelism via \textit{data parallel primitives} (DPPs), which enable
portable performance over different architectures.
Finally, since our algorithm works on MRFs in parallel and uses DPPs, we refer to the algorithm as DPP-PMRF.

Our results show that DPP-PMRF yields significant 
speedups.
We compared it to a serial implementation on a variety of platforms, and observed speedups of up to 13X (CPU) and 44X (GPU).
We also developed a reference algorithm that uses OpenMP to parallelize the MRF graph optimization in a coarsely-parallel manner.
We found that our DPP-based approach was up to 7X faster than the reference algorithm (CPU).
%
These performance gains are a result of the fine-grained parallel design of DPP-PMRF, which reformulates the MRF graph optimization in terms of multiple data-parallel operations over 1D arrays.
This DPP-based design is particularly amenable for the available instruction vectorization and high arithmetic and memory throughput inherent in modern CPU and GPU architectures.  

The contribution of this paper is two-fold.
First, we have developed a novel, DPP-based, shared-memory parallel implementation of an algorithm for a graph-labeling, data analysis problem.
This contribution is important for our science objectives, since it helps to improve platform-portable performance and data throughput of our overall image analysis workflow.
Second, we provide additional evidence that the DPP approach ---already quite popular for scientific visualization--- also works on graph-based data analysis problems.

\S\ref{sec:background} gives an overview about probabilistic graphical models applied to image segmentation. 
Additionally, we present some related works on performance and portability in graph-based methods and using data parallel primitives.
\S\ref{sec:design} presents our implementation, with particular attention to how an existing implementation of the MRF-based image segmentation method is expressed using DPPs.
\S\ref{sec:results} evaluates the DPP-based implementation in terms of correctness of results, and in terms of strong scaling studies on a multi-core platform, which includes a comparison with a reference OpenMP-based implementation.
\section{Background and Related Work}
\label{sec:background}

\subsection{MRF-based Image Segmentation}
Image segmentation is a compute-intensive task, and is a key component of multi-stage scientific analysis pipelines,
particularly those that work with large-scale image-based data obtained by experiments and advanced instruments, such as the X-ray imaging devices located at the Advanced Light Source at Berkeley Lab (Advanced Light Source website: \url{http://als.lbl.gov/}).  
As such instruments continually update in spatial and spectral resolution, there is an increasing need for high-throughput processing of large collections of 2D and 3D image data for use in time-critical activities such as experiment optimization and tuning~\cite{Bethel:eScience:2016}.
Our work here is motivated by the need for image analysis tools that perform well on modern platforms, and that are expected to be portable to next-generation hardware.

The process of segmenting an image involves separating various phases or components from the picture using photometric information and/or relationships between pixels/regions representing a scene. 
This essential step in an image analysis pipeline has been given great attention recently when studying experimental data~\cite{Perciano:JSR17}. There are several different types of image segmentation algorithms, which can be divided into categories such as: threshold-based, region-based, edge-based, clustering-based, graph-based and learning-based techniques. Of these, the graph- and learning-based methods tend to present the highest accuracy, but also the highest computational cost.


Graph-based methods are well-suited for image segmentation tasks due to their ability to use contextual information contained in the image, i.e., relationships among pixels and/or regions. 
The probabilistic graphical model (PGM) known as Markov random fields (MRF)~\cite{li:mrfmodel:2013} is an example of one such method. MRFs represent discrete data by modeling neighborhood relationships, thereby consolidating structure representation for image analysis~\cite{lezoray:graphs:2012}.

An image segmentation problem can be formulated using an MRF model on a graph $G$, where the segmented image is obtained through an optimization process to find the best labeling of the graph.
The graph $G(V,E)$ is constructed from an input image, where $V$ is the set of nodes and $E$ is the set of edges. Each node $V_i$ represents a region (set of pixels) and two nodes, $V_i$ and $V_j$, are connected by an edge if their corresponding regions share a boundary.

In an MRF model, the optimization process uses a global energy function to find the best solution to a similarity problem, such as the best pixel space partition. This energy function consists of a
data term and a smoothness term. 
For image segmentation, we use the mean of the intensity values of a region as the data term.
The smoothness term takes into account the similarity
between regions. 
The goal is to find the best labeling for the regions, so that the similarity between two regions with the same labels is optimal for all pixels~\cite{Mahapatra:2012}.

Given an image represented by $\mathbf{y} = (y_1,\dots,y_N)$, where each $y_i$ is a region, we want a configuration of labels $\mathbf{x} = (x_1,\dots,x_N)$ where $x_i \in L$ and $L$ is the set of all possible labels, $L = \{0, 1, 2,\dots, M\}$. 
The MAP criterion~\cite{li:mrfmodel:2013} states that one wants to find a labeling $\mathbf{x}^*$ 
that satisfies $\mathbf{x}^* = \underset{x}{\operatorname{argmax}}\{P(\mathbf{y}|\mathbf{x},\Theta)P(\mathbf{x})\}$,
which can be rewritten in terms of the energies~\cite{li:mrfmodel:2013} as
$\mathbf{x}^* = \underset{x}{\operatorname{argmin}}\{U(\mathbf{y}|\mathbf{x},\Theta) + U(\mathbf{x})\}$ (please refer to ~\cite{Perciano:ICIP16} for details regarding the prior and likelihood energies used in our approach).

Despite their high accuracy, MRF optimization algorithms have high computational complexity (NP-hard).
Strategies for overcoming the complexity, such as graph-cut techniques,
are often restricted to specific types of models (first-order MRFs)~\cite{1262177} and energy functions (regular or submodular)~\cite{1262177}.
%
%
In order to circumvent such drawbacks, recent works~\cite{Meng:2013, meng:likelihoods:2014} have proposed theoretical foundations for distributed parameter estimation in MRF. 
These approaches make use of a composite likelihood, which enables parallel solutions to sub problems. 
Under general conditions on the composite likelihood factorizations, the distributed estimators are proven to be consistent. 
The Linear and Parallel (LAP)~\cite{mizrahi:lap:2014} algorithm parallelizes naturally over cliques and, for graphs of bounded degree, its complexity is linear in the number of cliques. 
It is fully parallel and, for log-linear models, it is also data efficient. 
It requires only the local statistics of the data, i.e., considering only pixel values of local neighborhoods, to estimate parameters.

Perciano \textit{et al.}~\cite{Perciano:ICIP16} describe a graph-based model, referred to as Parallel Markov random fields (PMRF), which exploits MRFs to segment images. Both the optimization and parameter estimation processes are parallelized using the LAP method. 
In the work we present here, we use an OpenMP-based PMRF implementation as the ``reference implementation,'' and reformulate this method using DPPs. We study the viability of using DPPs as an alternative way to formulate an implementation to this challenging graph-based optimization problem, and compare shared-memory scalability of the DPP and reference implementation.

\subsection{Performance and Portability in Graph-based Methods}


The idea of formulating algorithms as sequences of highly optimized kernels, or motifs, is not new: this approach has formed the basis for nearly all numerical library and high performance simulation work going back almost 40 years, to the early implementations of LINPACK~\cite{CPE:CPE728}.
Over the years, several different highly optimized and parallel-capable linear algebra libraries have emerged, which serve as the basis for constructing a diverse collection of scientific computing applications. 
Such libraries include ScaLAPACK~\cite{choi1996scalapack}, BLASFEO (Basic Linear Algebra Subroutines for Embedded Optimization)~\cite{DBLP:journals/corr/FrisonKZD17} and MAGMA (Matrix Algebra on GPU and Multicore Architectures)~\cite{magma:2010}, to name a few.
%

The concept of using combinations of highly optimized building blocks has served as guiding design principle for many works focusing on high performance graph processing tools.
The Boost Graph Library (BGL)~\cite{siek2002boost} is a seminal implementation of data structures and methods for operating on graphs. 
The Multi-thread Graph Library (MTGL)~\cite{Berry:MTGL:2007} adapts and focuses BGL design principles for use on multithreaded architectures, where latencies associated with irregular memory access are accommodated by increasing the thread count to fully utilize memory bandwidth.
More recent works, such as CombBLAS~\cite{Buluc:2011:CBD:2076556.2076566} and GraphBLAS~\cite{Buluc:GraphBLAS:2011,Kepner:GraphBLAS:2016}, provide the means to implement graph-based algorithms as sequences of linear algebra operations, with special attention to the irregular access patterns of sparse vector and matrix operations, and on distributed-memory platforms.
GraphMat~\cite{Sundaram:2015:GHP:2809974.2809983} provides the means to write vertex programs and map them to generalized sparse matrix vector multiplication operations that are highly optimized for multi-core processors.
The STAPL parallel graph library~\cite{Harshvardhan2013} focuses more on the data structures and infrastructure for supporting distributed computations that implement computational patterns (e.g., map-reduce) for user-written graph algorithms.
%


Like many of these previous works, we are also examining the concept of platform portable graph algorithm construction using optimized building blocks. Compared to these previous works, our focus is narrower in terms of graph algorithm (PGM optimization) and building block (DPP).

\subsection{Performance and Portability with Data Parallel Primitives}
\label{sec:background:dpp}

The primary motivation for focusing on data parallel methods, particularly those that are amenable to vectorization, is because this approach appears promising for achieving good performance on multi- and many-core architectures, where there is an increasing on-chip computational capacity but relatively flat growth in memory bandwidth.
Levesque and Voss~\cite{MPP-Programming:2017} speculate that vectorized codes may achieve performance gains of as much as 10-30 fold compared to non-vectorized code, with the added benefit of using less power on multi- and many-core architectures.
%
%
DPPs are amenable to vectorization, and in turn, are capable 
of high performance on multi- and many-core architectures. 
This idea is not new, but dates back over 20 years to work by Blelloch~\cite{Blelloch:1990}, who proposed a vector model for parallel computing.

The following are examples of canonical DPPs that are used as building blocks to construct data-parallel algorithms:
\begin{tightItemize}
\item \textit{Map}: Invokes the same operation on each element of the input array, storing the result in the corresponding location of an output array of the same size;
\item \textit{Reduce}: Applies a binary operation (e.g., minimum or summation) on all elements of an input array, returning a single aggregate output value. 
ific\textit{ReduceByKey}: Performs a segmented Reduce on the input array, with segments based on unique keys, or data values, yielding an aggregate output value for each unique key;
\item \textit{Scan}: Calculates a series of partial summations, or a prefix sum, over the data values in an input array, producing an output array of the same size;
\item \textit{Scatter}: Writes each value of an input data array into a location in an output array, as specified in an input array of write indices;
\item \textit{SortByKey} Conducts an in-place segmented Sort on the input array, with segments based on unique keys, or data values, in the input array;
\item \textit{Unique}: Ignores duplicate values which are adjacent to each other, copying only unique values from the input array to the output array of the same or lesser size.
\end{tightItemize}

For this study, we have reformulated the MRF optimization problem entirely in terms of DPPs that are implemented as part of the VTK-m library of data analysis and visualization algorithms for emerging processor architectures~\cite{vtk-m,Moreland:CGA2016}.
VTK-m is a platform-portable framework that provides a set of key DPPs, along with back-end code generation and runtime support for use on GPUs (NVIDIA CUDA Toolkit~\cite{Nvidia-ProgramGuide}) and multi-core CPUs (Intel Thread Building Blocks (TBB)~\cite{TBB:web:2017}), all from a single code base. 

Specifically, DPPs are called via high-level function names specified by VTK-m.
The underlying data-parallel operations are then executed via low-level function calls of a platform-specific library, such as TBB for CPUs and NVIDIA Thrust~\cite{Thrust:web:2017} for GPUs. 
Thus, a single VTK-m code base written in terms of DPP can be executed across multiple platforms and methods of parallel processing.
Moreover, VTK-m code (and DPP-based design, in general) is robust to emerging platform architectures.
Given a new platform (e.g, FPGA) or multi-threading framework (e.g., OpenMP and OpenCL) for an existing platform, each DPP just needs to be implemented in terms of an optimized data-parallel library or code base native to the platform of execution (e.g., OpenMP-based DPP for CPUs or OpenGL-based DPP for GPUs).
In this case, the DPPs would still be called with the same high-level VTK-m function names, and invoke the underlying platform-specific library functions. 

%
VTK-m's primary focus thus far has been on platform portable scientific
visualization applications, with recent work showing viability in terms of portability, scalability, and performance gains within the context of ray-tracing~\cite{Larsen:PacVis2015}, unstructured volume  rendering~\cite{Larsen:EGPGV2015}, isocontouring~\cite{lo2012piston}, cell-projection volume rendering~\cite{Schroots15}, external facelist calculation~\cite{Lessley:EFC:EGPGV:2016}, and wavelet compression~\cite{Li:WC:EGPGV:2017}.

While VTK-m's use as a vehicle for achieving platform portability and performance for visualization methods is becoming better understood, its use as the basis for platform portable analysis computations is largely unexplored. 
Recent work~\cite{Lessley:DPP-MCE:LDAV:2017} uses a DPP formulation of a graph analytics problem, namely maximal clique enumeration (MCE).
The results show that the DPP reformulation is competitive with a state-of-the-art implementation in locating maximal cliques, is platform portable with performance analysis on both CPU and GPU platforms, and  offers significant evidence that this approach is viable for use on graph-based problems. 

An open question, which is outside the scope of this work, is whether or not the MRF optimization problem can be recast in a way that leverages platform-portable and parallel implementations such as GraphBLAS~\cite{Buluc:GraphBLAS:2011,Kepner:GraphBLAS:2016}, which accelerates graph operations by recasting computations as sparse linear algebra problems.
Unlike many graph problems, the MRF optimization problem here is not a sparse-data problem: as part of the problem setup, the graphical model is represented internally, in the reference implementation, in dense array form, and then the energy optimization computations are performed on densely packed arrays.
Our DPP-PMRF implementation recasts these dense-memory computations using DPPs, which are highly amenable to vectorization.
The primary focus of this study is to better understand the performance comparison and characterization between a reference implementation and one derived from DPPs. 


\section{Design and Implementation}
\label{sec:design}

This section introduces our new DPP-based PMRF image segmentation algorithm, which we refer to as DPP-PMRF.
We first review the foundational PMRF approach upon which our work is based, and then present our reformulation of this approach using DPP.

\subsection{The Parallel MRF Algorithm}
\label{sec:design:pmrf}

The parallel MRF algorithm (PMRF) proposed by Perciano et al.~\cite{Perciano:ICIP16} is shown in Algorithm~\ref{alg:pseudocode}. 
It consists of a one-time initialization phase, followed by a compute-intensive, primary parameter estimation optimization phase.
The output is a segmented image.

\begin{algorithm}[t]
\caption{Parallel MRF}\label{euclid}
\label{alg:pseudocode}
\footnotesize
\begin{algorithmic}[1]
\Require Original image, oversegmentation, number of output labels
\Ensure Segmented image and estimated parameters
\State Initialize parameters and labels randomly
\State Create graph from oversegmentation
\State Calculate and initialize $k$-neighborhoods from graph
\For{each EM iteration}
\For {each neighborhood of the subgraph} 
\State Compute MAP estimation
\State Update parameters 
\EndFor
\State Update labels
\EndFor
\end{algorithmic}
\end{algorithm}



The goal of the initialization phase is the construction of an undirected graph of pixel regions.
The graph is built based on an oversegmented version of the original input image.
An oversegmentation is a partition of the image into non-overlapping regions (superpixels), each with statistically similar grayscale intensities among member pixels~\cite{1335450}.
The partitioning of the image we are using in this work is irregular, i.e. the non-overlapping regions can have different sizes and shapes.
%
%
Each vertex $V$ of the graph represents a region in the oversegmented image (i.e., a spatially connected region of pixels having similar intensity), and each edge $E$ indicates spatial adjacency between regions. Given the irregular nature of the oversegmentation, the topological structure of the graph varies accordingly.
%

%
Next, in the main computational phase, we define an MRF model over the set of vertices, which includes an energy function representing contextual information of the image.
In particular, this model specifies a probability distribution over the $k$-neighborhoods of the graph.
Each $k$-neighborhood consists of the vertices of a maximal clique, along with all neighbor vertices that are within $k$ edges (or hops) from any of the clique vertices; in this study, we use $k=1$.
Using OpenMP, the PMRF algorithm performs energy function optimization, in parallel, over the neighborhoods, each of which is stored as a single row in a \textit{ragged array}.
This optimization consists of an iterative invocation of the expectation-maximization (EM) algorithm, which performs parameter estimation using the maximum \textit{a posteriori} (MAP) inference algorithm~\cite{Koller:PGM}.
The goal of the optimization routine is to converge on the most-likely (minimum-energy) assignment of labels for the vertices in the graph; the mapping of the vertex labels back to pixels yields the output image segmentation.

Our proposed DPP-based algorithm overcomes several important problems encountered in the PMRF implementation such as non-parallelized steps of the algorithm (e.g., partitioning of the graph and MAP estimation computation), and platform portability.
In particular, the OpenMP design of the PMRF conducts outer-parallelism over the MRF neighborhoods, but does not perform inner-parallelism of the optimization phase for each neighborhood (e.g., the energy function computations and parameter updates).
Thus, the ability to attain fine-grained concurrency and greater parallelism is limited by the non-parallel computations within each outer-parallel optimization task.
Finally, for the construction of MRF neighborhoods, our new method makes use of a recent work on maximal clique enumeration using DPPs~\cite{Lessley:DPP-MCE:LDAV:2017}.
%

\subsection{DPP Formulation of PMRF}
\label{sec:design:dpp-pmrf}

We now describe our DPP-based PMRF algorithm (DPP-PMRF) to perform image segmentation.
Our algorithm redesigns PMRF in terms of DPPs to realize outer-level parallelism over MRF neighborhoods, and inner-level parallelism within the optimization routine for the vertices of each neighborhood.
This data-parallel approach consists of an initialization phase followed by the main MRF optimization phase; refer to Algorithm~\ref{alg:dpp-pmrf-pseudocode} for the primary data-parallel steps.
%

\subsubsection{Initialization}
In this initial phase, we first construct an undirected graph $G$ representing the connectivity among oversegmented pixel regions in the input image; refer to \S\ref{sec:design:pmrf};
Then, we enumerate all of the maximal cliques within $G$, yielding a set of complete subgraphs that form the basis of the MRF neighborhood structure.

Our initialization procedure is similar to that of the reference PMRF, but differs in the following ways.
First, all of our initialization operations and algorithms are designed in terms of DPP, exposing high levels of data-parallelism throughout the entire image segmentation pipeline; refer to~\cite{Lessley:DPP-MCE:LDAV:2017} for the DPP-based enumeration of maximal cliques.
Second, we represent $G$ in a compressed, sparse row (CSR) format that fits compactly within shared memory; see ~\cite{Lessley:DPP-MCE:LDAV:2017} for details on the DPP construction of this graph.
%

\subsubsection{Optimization}

Given the graph $G$ and its set of maximal cliques from the initialization, we proceed to the optimization phase, which consists of the following two primary data-parallel tasks: construction of neighborhoods over the maximal cliques and EM parameter estimation, the latter of which comprises the main computational work in this phase.
Prior to constructing the neighborhoods, the mean and standard deviation parameters, $\mu$ and $\sigma$, of each label are randomly initialized to values between $0$ and $255$, representing the $8$-bit grayscale intensity spectrum; in this study we focus on binary image segmentation with two labels of $0$ and $1$.
Additionally, the label for each vertex of $G$ is randomly initialized to either $0$ or $1$.
\newline


\textbf{Construction of Neighborhoods}: In the PMRF algorithm, $1$-neighborhoods are serially constructed from maximal cliques during the initialization process of the algorithm.
%
%
Our approach constructs the $1$-neighborhoods before the parameter estimation phase and consists of the following data-parallel steps that operate on individual vertices, as opposed to entire maximal cliques, exposing more inner, fine-grained parallelism.
\begin{tightEnumerate}
\item \textbf{Find Neighbors}: Invoke a data-parallel \textit{Map} primitive to obtain, for each vertex, a count of the number of neighbors that are within 1 edge from the vertex and not a member of the vertex's maximal clique.
\item \textbf{Count Neighbors}: Call a \textit{Scan} primitive to add the neighbor counts, the sum of which is used to allocate a neighborhoods array.
\item \textbf{Get Neighbors}: In a second pass to a \textit{Map} primitive, populate the neighborhoods array with the neighbors, parallelizing over vertices as before.
\item \textbf{Remove Duplicate Neighbors}: Since multiple vertices within the same maximal clique may output common $1$-neighbors in the neighborhoods array, successively invoke \textit{SortByKey} and \textit{Unique} primitives to remove the duplicate neighbors. 
The \textit{SortByKey} primitive contiguously arranges vertices in the array in ascending order of their vertex Id and clique Id pairs. 
Then, the \textit{Unique} primitive removes these duplicate, adjacent vertices, leaving a final neighborhoods array in which each set of neighbors is arranged in order of vertex Id.
\end{tightEnumerate}

\begin{algorithm}[t]
\caption{DPP-PMRF}
\label{alg:dpp-pmrf-pseudocode}
\footnotesize
\begin{algorithmic}[1]
\Require Original image, oversegmentation, number of output labels
\Ensure Segmented image and estimated parameters
\State Create graph from oversegmentation in parallel
\State Enumerate maximal cliques of graph in parallel
\State Initialize parameters and labels randomly
\State Construct $k$-neighborhoods from maximal cliques in parallel
\State Replicate neighborhoods by label in parallel
\For{each EM iteration}
\State Gather replicated parameters and labels in parallel
\For {each vertex of each neighborhood} 
\State MAP estimation computed in parallel
\EndFor
\State Update labels and parameters in parallel
\EndFor
\end{algorithmic}
\end{algorithm}

\textbf{EM Parameter Estimation}: 
We formulate the EM parameter estimation via the following data-parallel steps.
\begin{tightEnumerate}
\item \textbf{Replicate Neighborhoods By Label}: Next, each neighborhood is replicated for each of the two class output labels.
With a sequence of \textit{Map}, \textit{Scan}, and \textit{Gather} DPPs, we obtain a set of expanded indexing arrays, each of size $2 \times |hoods|$.
The $testLabel$ array indicates which replication of the neighborhood a given element belongs to; e.g., vertex element 2 belongs to the first copy of its neighborhood, denoted by a 0 label.
The $hoodId$ array gives the Id of the neighborhood to which element belongs, and the $oldIndex$ array contains back-indices into the original $hoods$ array, for each replicated element.
\begin{align*}
hoods &= [0\; 1\; 2\; 5\; 1\; 3\; 4]\\
testLabel &= [0\; 0\; 0\; 0\; 1\; 1\; 1\; 1\; 0\; 0\; 0\; 1\; 1\; 1]\\
oldIndex &= [0\; 1\; 2\; 3\; 0\; 1\; 2\; 3\; 4\; 5\; 6\; 4\; 5\; 6]\\
hoodId &= [0\; 0\; 0\; 0\; 0\; 0\; 0\; 0\; 1\; 1\; 1\; 1\; 1\; 1]\\
repHoods &= [\underbrace{0\; 1\; 2\; 5}_{\substack{Hood_0\\ Label_0}}\; \underbrace{0\; 1\; 2\; 5}_{\substack{Hood_0\\Label_1}}\; 1\; 3\; 4\; 1\; 3\; 4]
\end{align*}
The replication of the $hoods$ array, $repHoods$, is not allocated in memory, but is simulated on-the-fly with a memory-free \textit{Gather} DPP using $oldIndex$.
\item \textbf{For each EM iteration $i$}:
       \begin{tightItemize}
       \item \textbf{Compute Energy Function}: Using the array of back-indices ($oldIndex$) into the neighborhoods array ($hoods$), we invoke a set of \textit{Gather} DPP to create replicated data arrays of size $2 \times |hoods|$:
\begin{align*}
vertLabel &= [1\; 1\; 0\; 1\; 1\; 1\; 0\; 1\; 1\; 0\; 1\; 1\; 0\; 1]\\
vertMu &= [40\; 20\; 55\; 25\; 40\; 20\; 55\; 25\; 20\; 65\; 35\; 20\; 65\; 35]\\
labelMu &= [30\; 30\; 60\; 30\; 30\; 30\; 60\; 30\; 30\; 60\; 30\; 30\; 60\; 30]
\end{align*}
        We then invoke a \textit{Map} DPP to compute an energy function value for each of the replicated neighborhood vertices.
        This operation parallelizes over the data arrays and calculates, for each vertex of a neighborhood, the energy, or deviation, between its actual grayscale intensity value ($vertMu$) and that of the label mean parameter ($labelMu$).
       \item \textbf{Compute Minimum Vertex and Label Energies}: Within the array of computed energy function values, each vertex of a given neighborhood is associated with two values, one for each of the labels.
       In order to determine the minimum energy value between these labels, we invoke a \textit{SortByKey} DPP, which makes each pair of energy values contiguous in memory.
       Then, we call consecutive \textit{ReduceByKey$\langle$Min$\rangle$} DPP on the sorted energy values to obtain the minimum energy value for each vertex:
       \item \textbf{Compute Neighborhood Energy Sums}: Given the minimum energies values, we call a \textit{ReduceByKey$\langle$Add$\rangle$} DPP to compute the sum of the values for each neighborhood.
       \item \textbf{MAP Convergence Check}: We maintain an array that stores the energy sum of each neighborhood at the end of every EM iteration.
       Using a \textit{Map} DPP, we measure the amount of change in neighborhood energy sums from the previous $L$ iterations ($L=3$ in this study), and mark a neighborhood as \textit{converged} if this change falls below a constant threshold of $1.0 \times 10^{-4}$.
       Once all neighborhoods have converged---assessed via a \textit{Scan} DPP primitive---we end the EM optimization.
       \end{tightItemize}
\item \textbf{Update Output Labels}: Invoke a \textit{Scatter} DPP to write the minimum-energy label of each neighborhood vertex to its corresponding location in the global vertex label array.
\item \textbf{Update Parameters}: Use a sequence of \textit{Map}, \textit{ReduceByKey}, \textit{Gather}, and \textit{Scatter} DPP, to update the parameters of each label ($\mu$ and $\sigma$) as a function of a) the intensity values of the vertices assigned to the labels (i.e., minimum energy labels), and b) the sum of the per-label and per-vertex energy function values.
\item \textbf{EM Convergence Check}: We maintain an array that stores, for each EM iteration, the total sum of the neighborhood energy value sums after the final MAP iteration.
Calling a \textit{Scan} DPP on these neighborhood sums yields this total EM sum.
Similar to the MAP convergence check, we assess, via a \textit{Map} DPP, the variation in EM sums over the previous $L$ iterations.
\end{tightEnumerate}
During our experimentation, we find that most invocations of the EM optimization converge within $20$ iterations; thus, we use that number of iterations in this study.
Finally, we return the estimated parameters and assignment of labels to vertices as output.
These labels can be mapped back to pixel regions of the vertices to produce the final segmented image.

\section{Results}
\label{sec:results}

The experimental results in this section serve to answer two primary questions. 
First, in \S\ref{sec:results:verification}, we examine the question of correctness: is the new DPP-PMRF algorithm producing correct results? 
Second, in \S\ref{sec:results:performanceScalability},  we are interested in understanding how well the DPP-PMRF implementation performs on different modern CPU and GPU platforms: does DPP-PMRF demonstrate platform-portable performance?
Because these experiments examine different questions, each uses a different methodology, which we present in conjunction with the experiment results.
\S\ref{sec:results:sourceDataPlatforms} describes the source datasets and the computational platforms that we use in both sets of experiments.

\subsection{Source Data, Reference Implementation, and Computational Platforms}
\label{sec:results:sourceDataPlatforms}

\subsubsection{Datasets}


We test the DPP-PMRF implementation using two types of image-based datasets: one is synthetic and the other is output from a scientific experiment.
The former is used to verify the accuracy of the proposed algorithm against a known benchmark that offers a ground-truth basis of comparison. The latter shows how DPP-PMRF performs on a real-world problem. 


\textit{Synthetic data.} 
We selected the synthetic dataset from the 3D benchmark made available by the Network Generation Comparison Forum (NGCF)~\footnote{\url{http://people.physics.anu.edu.au/~aps110/network_comparison}}. 
The NGCF datasets are a global, recognized standard to support the study of 3D tomographic data of porous media. The datasets provided are binary representations of a 3D porous media. For the purposes of this analysis, we corrupted the original stack by noise (salt-and-pepper) and additive Gaussian with $\sigma=100$. 
Additionally, we also simulate ringing artifacts~\cite{Perciano:JSR17} into the sample to closer resemble real-world results. For the segmentation algorithm analysis, the corrupted data serves as the ``original data'' and the binary stack as the ground-truth. A full synthetic dataset is $268$ MB in size, and consists of 512 image slices of dimensions $512\times 512$. The chosen dataset emulates a very porous fossiliferous outcrop carbonate, namely Mt. Gambier limestone from South Australia. Because of the more homogeneous characteristic of this dataset, its related graph contains a larger number of smaller-sized neighborhoods.

\textit{Experimental data.}
This dataset contains cross-sections of a geological sample and conveys information regarding the x-ray attenuation and density of the scanned material as a gray scale value. This data was generated by the Lawrence Berkeley National Laboratory Advanced Light Source X-ray beamline 8.3.2~\footnote{\url{microct.lbl.gov}}~\cite{Donatelli:2015}. The scanned samples are pre-processed using a separate software that provides reconstruction of the parallel beam projection data into a 3 GB stack of 500 image slices with dimensions of $1813\times 1830$.
This dataset contains a very different and more complex set of structures to be segmented. Consequently, compared to the synthetic data, this experimental data leads to a denser graph with many more neighborhoods of higher complexity.

\subsubsection{Hardware Platforms}
\label{sec:results:hardware}
Our verification and performance tests were run on two different multi-core platforms maintained by the National Energy Research Scientific Computing Center (NERSC). For each platform, all tests were run on a single node (among many available). Specifications for these two platforms are as follows:
\begin{tightEnumerate}
\item \texttt{Cori.nersc.gov} (KNL): Cray XC40 system with a partition of 
9,688 nodes, each containing a single-socket 68-core 1.4 GHz Intel Xeon Phi 7250 (Knights Landing (KNL)) processor and 96 GB DDR4 2400 GHz memory. With hyper-threading, each node contains a total of 272 logical cores (4 hyper-threads per core)\footnote{Cori configuration page: \url{http://www.nersc.gov/users/computational-systems/cori/configuration/}}.
\item \texttt{Edison.nersc.gov} (Edison): Cray XC30 system comprised of 5586 nodes, each containing two 12-core 2.4 GHz Intel Ivy Bridge processors (two sockets) and 64 GB DDR3 1866 MHz memory. With hyper-threading, each node contains a total of 48 logical cores (24 logical cores per socket of a node)\footnote{Edison configuration page: \url{http://www.nersc.gov/users/computational-systems/edison/configuration/}}
\end{tightEnumerate}
The intention of running on the KNL and Edison systems is to create an opportunity for revealing architecture-specific performance characteristics.

Performance tests were also conducted on a general-purpose GPU platform: 
\begin{tightEnumerate}
\item \texttt{K40}: NVIDIA Tesla K40 Accelerator with 2880 processor cores, 12 GB memory, and 288 GB/sec memory bandwidth. 
Each core has a base frequency of 745 MHz, while the GDDR5 memory runs at a base frequency of 3 GHz.
\end{tightEnumerate}

For both the experimental and synthetic image datasets, the peak memory usage of DPP-PMRF is well within the maximum available memory of the tested CPU (between 64 GB and 96 GB) and GPU (12 GB) platforms.
The execution of DPP-PMRF results in a maximum memory footprint of between 300 MB to 2 GB for the experimental images and between 100 MB and 400 MB for the synthetic images.
%

\begin{figure*}[!th]
\centering
\subfloat[Original]{\includegraphics[width=0.22\linewidth]{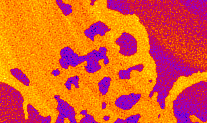}}\quad
\subfloat[Ground-truth]{\includegraphics[width=0.22\linewidth]{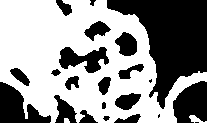}}\quad
\subfloat[DPP-PMRF result]{\includegraphics[width=0.22\linewidth]{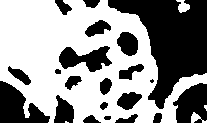}}\quad
\subfloat[Simple threshold]{\includegraphics[width=0.22\linewidth]{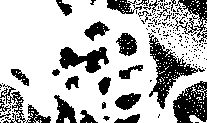}}\\
\subfloat[3D rendering of the original data]{\includegraphics[width=0.19\linewidth]{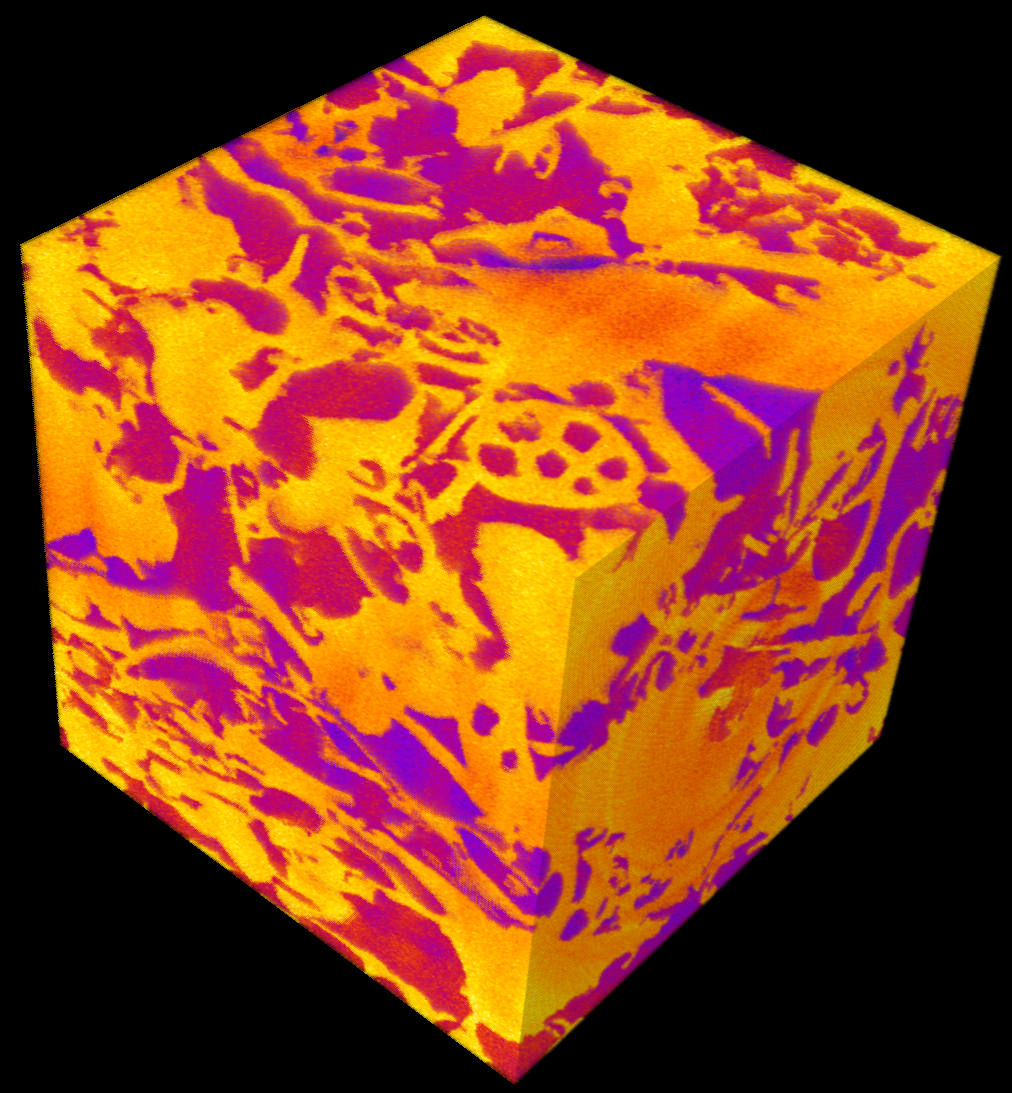}}\quad
\subfloat[3D rendering of the DPP-PMRF result]{\includegraphics[width=0.19\linewidth]{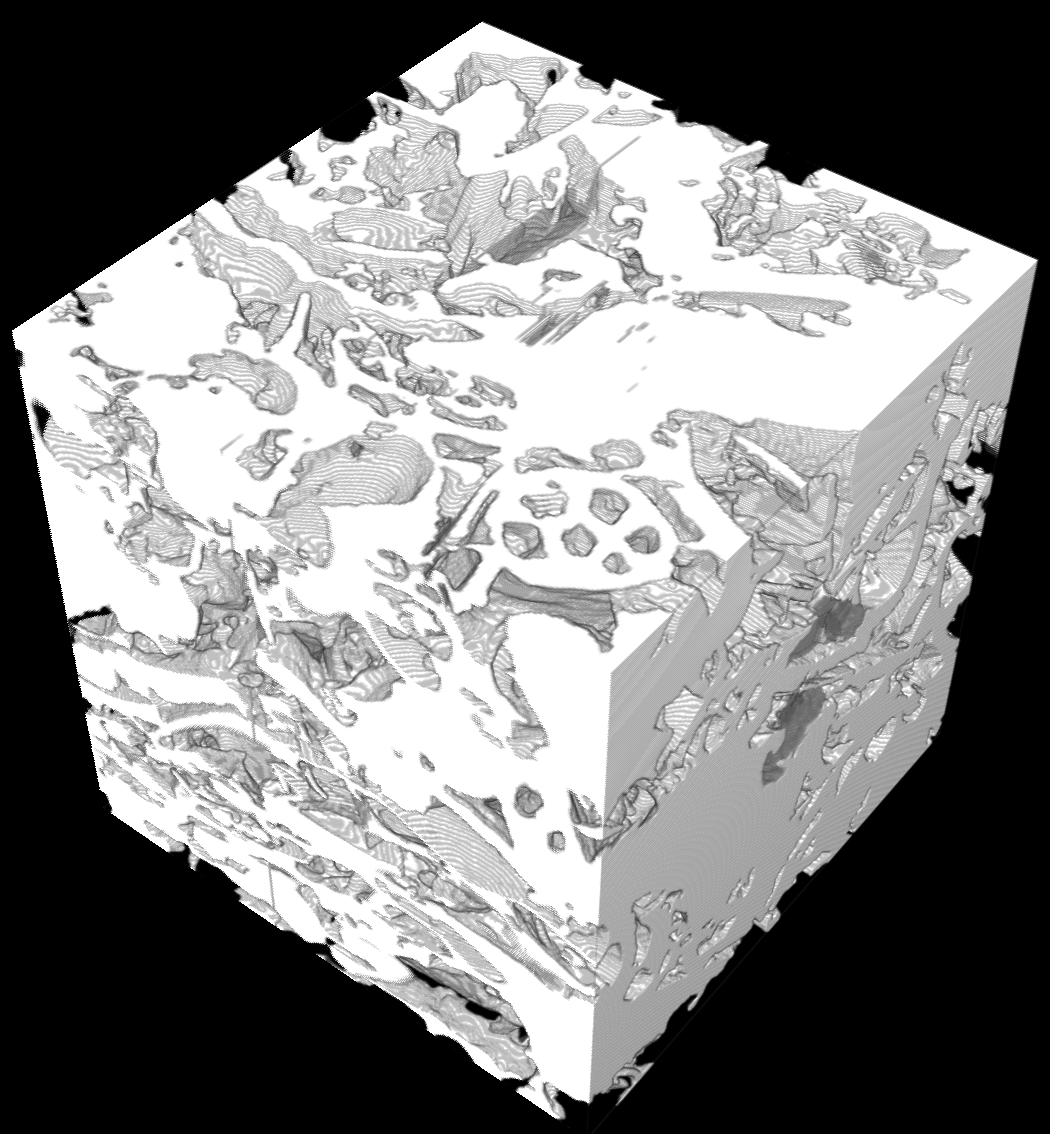}}
\caption{Results applying DPP-PMRF to the synthetic dataset. (a) Region of interest from the noisy data; (b) Ground-truth; (c) Result obtained by the proposed DPP-PMRF; (d) Result obtained using a simple threshold; (e) 3D rendering of the original noisy dataset; (f) 3D rendering of the result obtained by DPP-PMRF.
}
\centering
\label{fig:threedimsyn}
\end{figure*}

\begin{figure*}[!ht]
\centering
\subfloat[Original]{\includegraphics[width=0.2\linewidth]{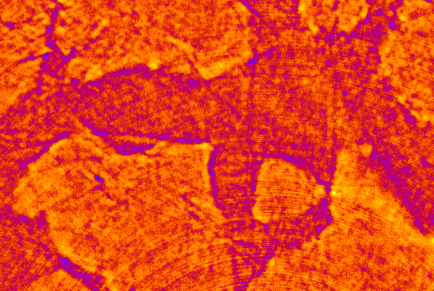}}\quad
\subfloat[Reference result]{\includegraphics[width=0.2\linewidth]{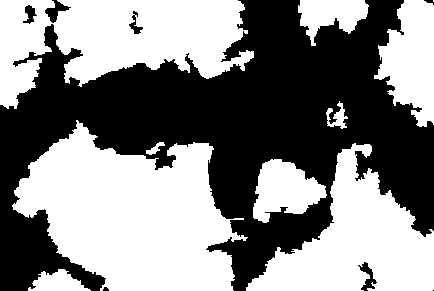}}\quad
\subfloat[DPP-PMRF result]{\includegraphics[width=0.2\linewidth]{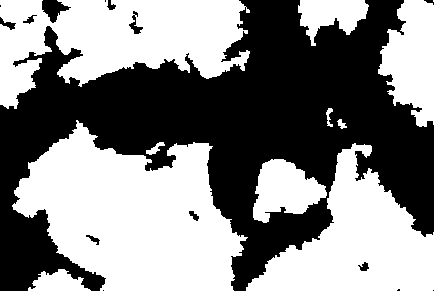}}\quad
\subfloat[Simple threshold]{\includegraphics[width=0.2\linewidth]{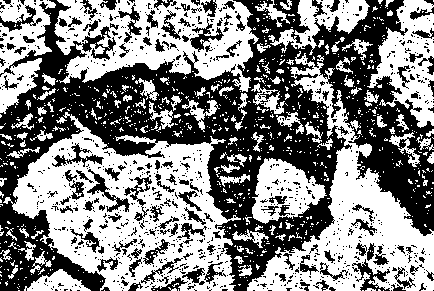}}\\
\subfloat[3D rendering of the original data]{\includegraphics[width=0.206\linewidth]{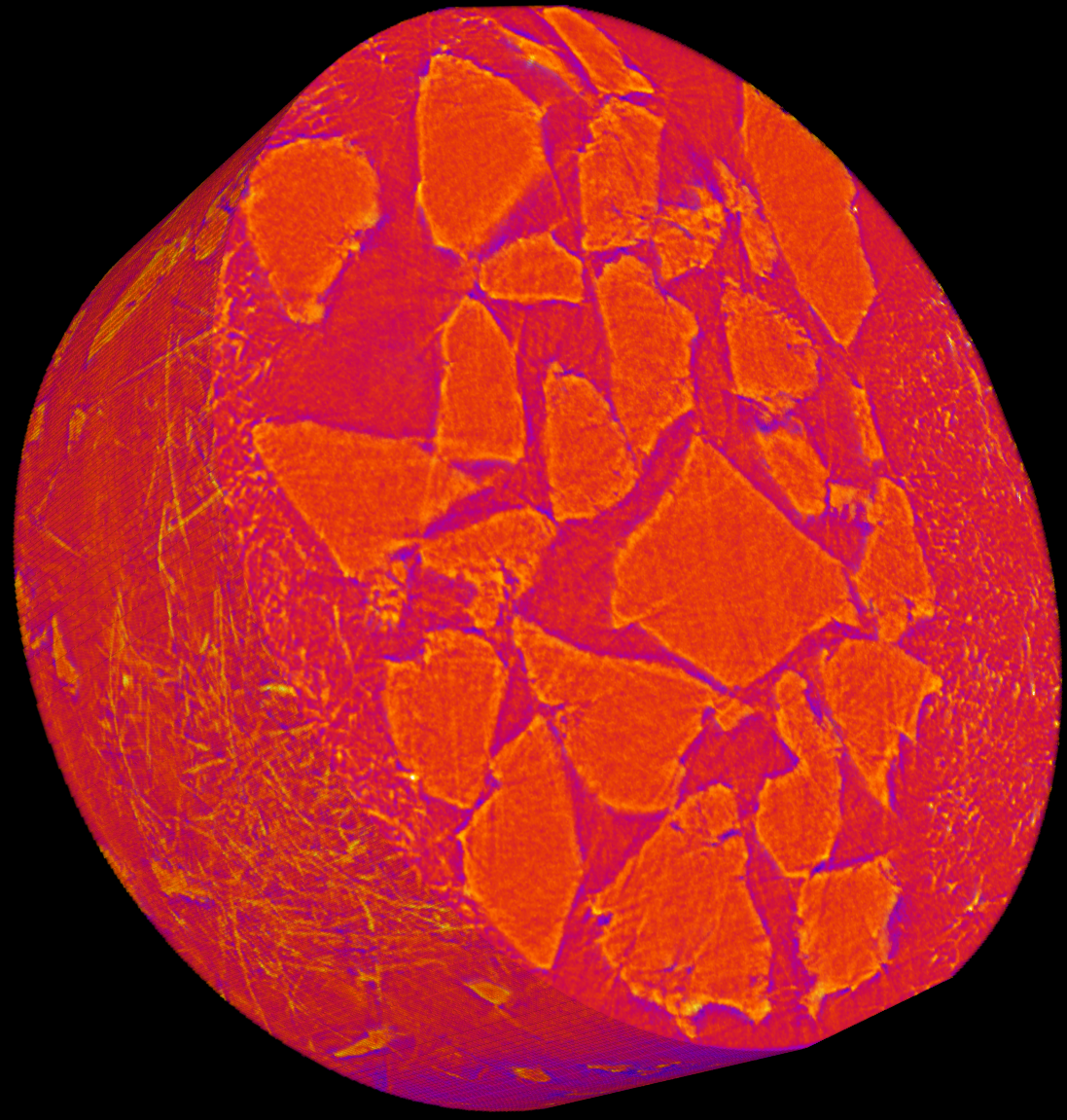}}\quad
\subfloat[3D rendering of the DPP-PMRF result]{\includegraphics[width=0.21015\linewidth]{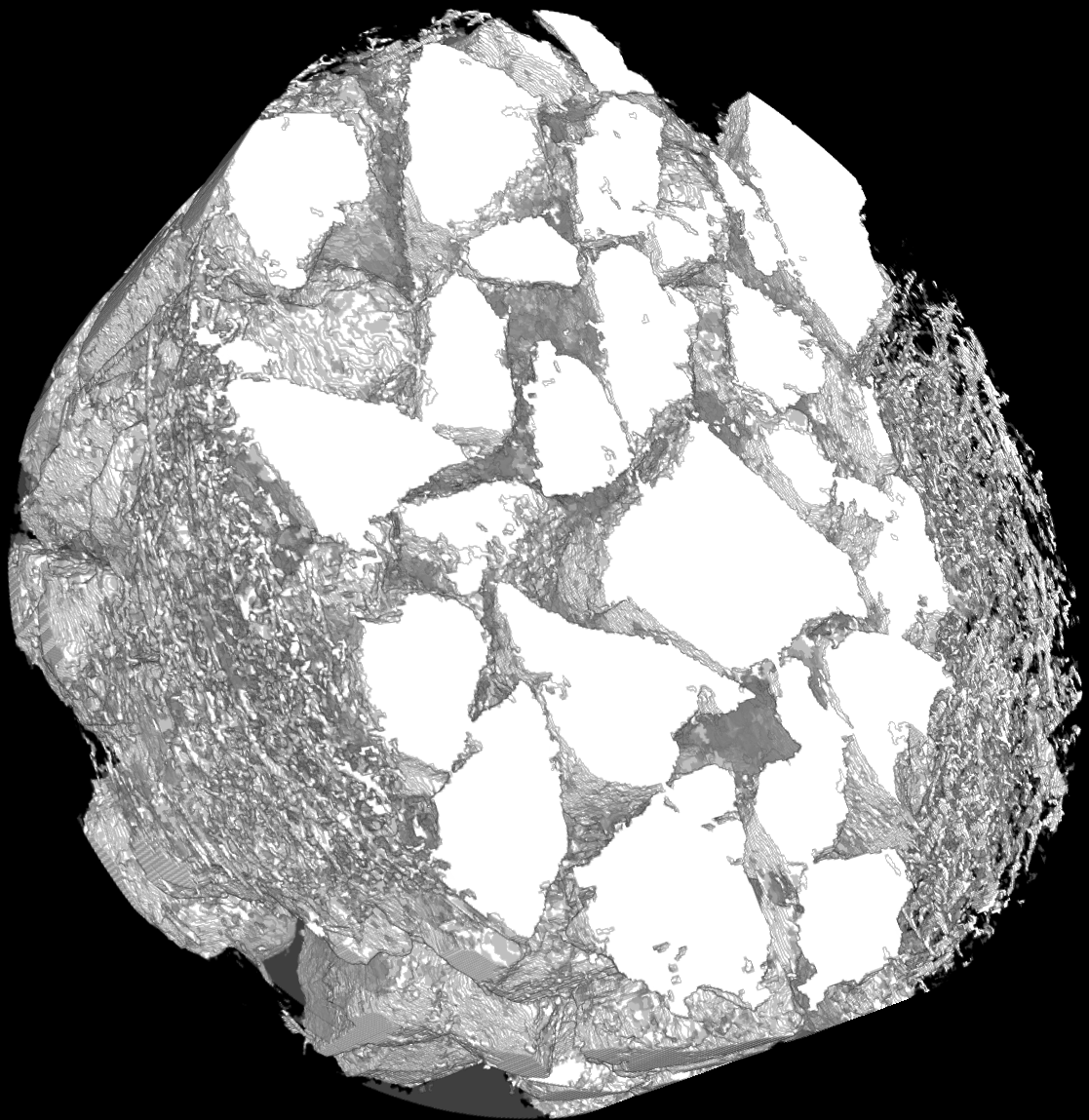}}
\caption{Results applying DPP-PMRF to the experimental dataset. (a) Region of interest from the original data; (b) Reference result; (c) Result obtained by the proposed DPP-PMRF; (d) Result obtained using a simple threshold; (e) 3D rendering of the original dataset; (f) 3D rendering of the result obtained by DPP-PMRF.
}
\centering
\label{fig:threedim}
\end{figure*}

\subsubsection{Software Environment}
\label{sec:results:software}

Our DPP-PMRF algorithm is implemented using the platform-portable VTK-m toolkit~\cite{vtk-m}.
With VTK-m, a developer chooses the DPPs to employ, and then customizes those primitives with functors of C++-compliant code.
This code then invokes back-end, architecture-specific code for the architecture of execution (enabled at compile-time), e.g., CUDA Thrust code for NVIDIA GPUs and Threading Building Blocks (TBB) code for Intel CPUs. 

In our CPU-based experiments, VTK-m was compiled with TBB enabled (version $17.0.2.174$) using the following C++11 compilers: GNU GCC $7.1.0$ on Edison and Intel ICC $18.0.1$ on KNL. In our GPU-based experiments, VTK-m was compiled with CUDA enabled (version $8.0.61$) using the NVIDIA CUDA NVCC compiler. For all experiments, version $1.2.0$ of VTK-m was used. 

With TBB enabled in VTK-m, each invocation of a DPP executes the underlying TBB parallel algorithm implementation for the primitive.
The basic input to each of these parallel algorithms is a linear array of data elements, a functor specifying the DPP operation, and a \textit{task} size that sets the number of contiguous elements a single thread can operate on. A partitioning unit invokes threads to recursively split, or divides in half, the array into smaller and smaller \textit{chunks}. During a split of a chunk, the splitting thread remains assigned to the left segment, while another \textit{ready} thread is assigned to the right segment. When a thread obtains a chunk the size of a task, it executes the DPP functor operation on the elements of the chunk and writes the results into an output array. Then, the thread is ready to be scheduled, or re-assigned, to another chunk that needs to be split further. This work-stealing scheduling procedure is designed to improve load balancing among parallel threads, while minimizing cache misses and cross-thread communication.  

\subsubsection{Reference Implementation of PMRF}
\label{sec:results:software:openmp}
In this study, we compare the performance and correctness of the new DPP-PMRF implementation with the PMRF reference implementation developed with OpenMP 4.5, which is described in \S\ref{sec:design:pmrf}.
We take advantage of OpenMP loop parallelism constructs to achieve outer-parallelism over MRF neighborhoods, and make use of OpenMP's dynamic  scheduling algorithm in the performance studies (see \S\ref{sec:results:scalability} for details).



The OpenMP-based implementation was built with the following C++11 compilers (same as for the DPP-based version): GNU GCC $7.1.0$ on Edison and Intel ICC $18.0.1$ on KNL. 
 
\subsection{Verification of Correctness}
\label{sec:results:verification}

The following subsections present a set of tests aimed at verifying that DPP-PMRF computes the correct, ground-truth image segmentation output.


\subsubsection{Methodology: Evaluation Metrics}

In order to determine the precision of the segmentation results we use the metrics $precision=\frac{TP}{TP+FP}$, $recall=\frac{TP}{TP + FN}$, and $accuracy=\frac{TP + TN}{TP + TN + FP + FN}$, where $TP$ stands for True Positives, $TN$ for True Negatives, $FP$ for False Positives, and $FN$ for False Negatives.

In addition, we also use the porosity (ratio between void space and total volume), or $\rho=\frac{V_v}{V_t},$, where $V_v$ is the volume of the void space and $V_t$ is the total volume of the void space and solid material combined.

\subsubsection{Verification Results}
\label{sec:verification}

Figure~\ref{fig:threedimsyn} shows the results of applying DPP-PMRF to the synthetic data. Figure~\ref{fig:threedimsyn}(a) presents a 2D region of interest from the corrupted data, Figures~\ref{fig:threedimsyn}(b-d) show the ground-truth, the result from DPP-PMRF and the result using a simple threshold, respectively, and Figures~\ref{fig:threedimsyn}(e-f) shows the 3D renderings of both the corrupted data and the DPP-PMRF result. We observe a high similarity between the DPP-PMRF result and the ground-truth, indicating a high precision when compared to the simple threshold result.
For this synthetic dataset, the verification metrics obtained are a precision of $99.3\%$, a recall of $98.3\%$, and an accuracy of $98.6\%$.


Following the same methodology, we present the results using the experimental dataset in Figure~\ref{fig:threedim}. Figures~\ref{fig:threedim}(a-d) shows regions of interest from the original data, the result from the reference implementation, the DPP-PMRF result, and a simple threshold result, respectively. Much like the results using the synthetic data, we observe a close similarity between the DPP-PMRF result and the reference result. The differences observed between the results are usually among the very small regions in the image, where small variations of labeling of the graph could lead to the same minimum energy value.
For this dataset the verification metrics obtained are a precision of $97.2\%$, a recall of $95.2\%$ and an accuracy of $96.8\%$.



\begin{figure*}[ht!]
\centering
\subfloat[Edison: Ivy Bridge]{\includegraphics[width=0.45\linewidth]{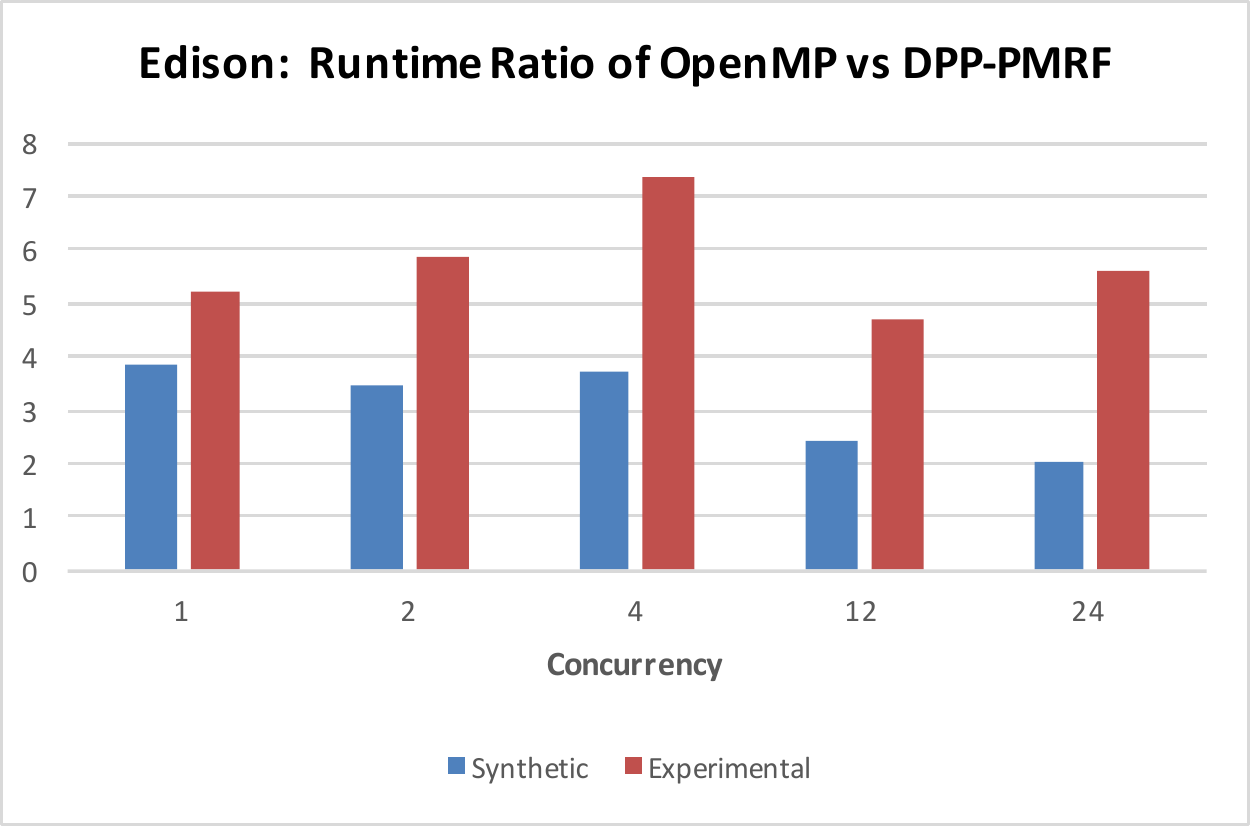}}
\hspace{2mm}
\subfloat[Cori: Knights Landing (KNL)]
{\includegraphics[width=0.45\linewidth]{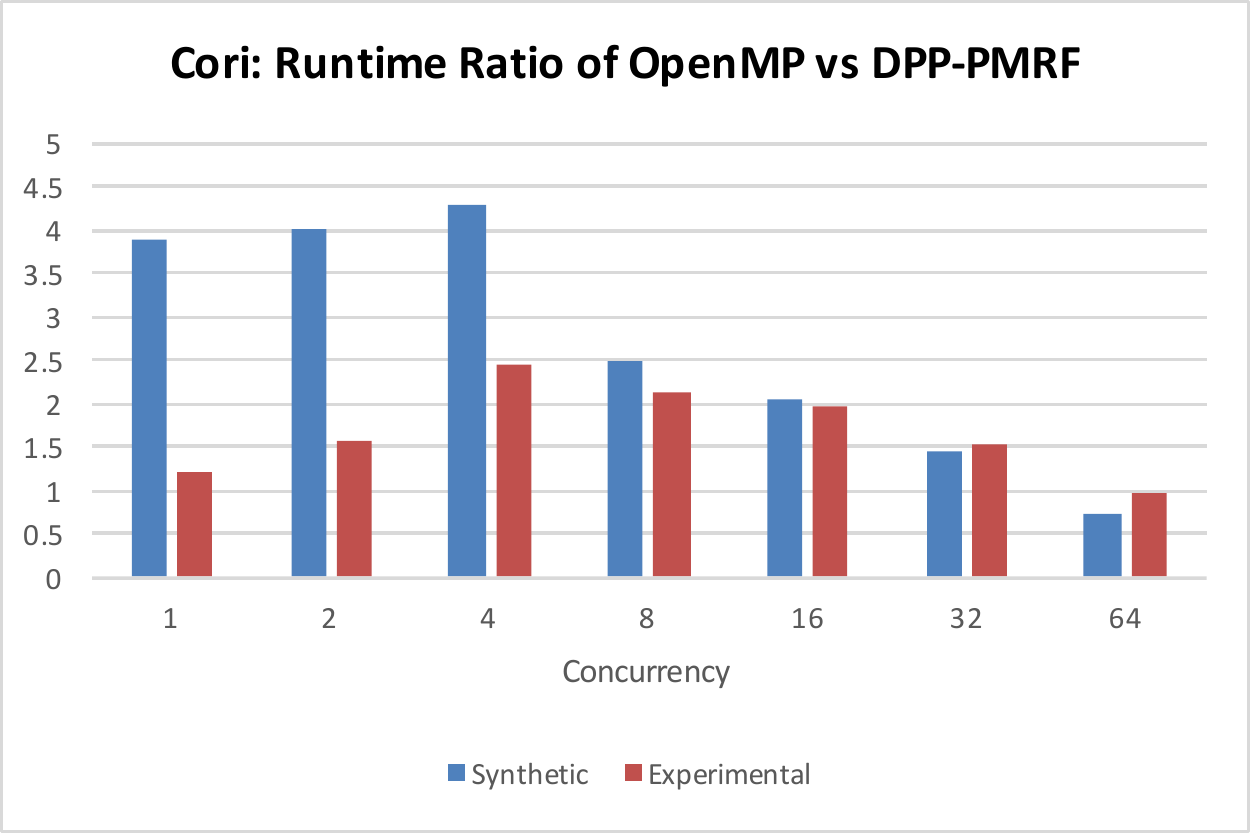}}
\caption{Comparison of absolute runtime of the DPP and OpenMP implementations at varying concurrency, on both platforms, and both sample datasets. The horizontal axis is concurrency level, or the number of physical cores used on a single node. Each bar represents the ratio of runtimes of the DPP-PMRF to the OpenMP code. The vertical axis measures how much faster the DPP-PMRF code is than the OpenMP code for a given dataset, on a given platform, and a given concurrency. A bar height of 1.0 means both codes have the same runtime; a bar height of 2.0 means the DPP code ran in half the time of the OpenMP code. See \S\ref{sec:results:runtime} for more details.}
\label{fig:runtimeComparison}
\end{figure*}

\subsection{Performance and Scalability Studies}
\label{sec:results:performanceScalability}


The next subsections present a set of studies aimed at verifying the performance, scalability, and platform-portability of DPP-PMRF, as compared to a OpenMP-parallel reference implementation and serial baseline.

\subsubsection{Methodology}


The objectives for our performance study are as follows. 
First, we are interested in comparing the absolute runtime performance of the OpenMP and DPP-PMRF shared-memory parallel implementations on modern multi-core CPU platforms.
Second, we wish to compare and contrast their scalability characteristics, and do so using a strong-scaling study, where we hold the problem size constant and increase concurrency.
Finally, we assess the platform-portable performance of DPP-PMRF by executing the algorithm on a general-purpose GPU platform and comparing the runtime performance to a serial (non-parallel) baseline and the CPU execution from the strong-scaling study.

To obtain elapsed runtime, we run these two implementations in a way where we iterate over 2D images of each 3D volume (synthetic data, experimental data).
We report a single runtime number, which is the average of elapsed runtime for each 2D image in the 3D volume. The runtime takes into account only the optimization process of the algorithm as this is the portion of the algorithm that is most computationally intensive.



From runtime measurements, we report results using charts that show time vs. concurrency, which are often known as "speedup" charts. Moreland and Oldfield \cite{moreland:metrics:2015} suggest that
such traditional performance metrics may not be effective for large-scale studies for problem sizes  that cannot fit on a single node. 
Since our problems all fit within a single node, and are of modest scale, we are showing traditional speedup charts.




Speedup is defined as $S(n, p) = \frac{T^* (n)}{T(n, p)}$ where \textit{T(n,p)} is the time it takes to run the parallel algorithm on \textit{p} processes with an input size of \textit{n}, and \textit{T*(n)} is the time for the best serial algorithm on the same input.





\subsubsection{CPU Runtime Comparison: OpenMP vs. DPP-PMRF}
\label{sec:results:runtime}

The first performance study question we examine is a comparison of runtimes between the OpenMP and DPP-PMRF implementations. 
We executed both codes at varying levels of concurrency on the Cori and Edison CPU platforms, using the two different datasets as input. 
Each concurrency level represents the number of physical cores used within a single node.
Hyper-threading was active in each experiment, resulting in more logical (virtual) cores than physical cores being utilized per node (see \S\ref{sec:results:hardware} for hardware configuration details).  
The runtimes for this battery of tests are presented in Fig.~\ref{fig:runtimeComparison} in a way that is intended to show the degree to which DPP-PMRF is faster, or slower, than the OpenMP version.

In Fig.~\ref{fig:runtimeComparison}, each bar is computed as the quotient of the OpenMP runtime and the DPP runtime. A bar height of 1.0 means both codes have the same runtime; a bar height of 2.0 means DPP-PMRF ran in half the time of the OpenMP code. A bar height of less than 1.0 would mean that the OpenMP code ran faster than DPP-PMRF.
%
%
%
These results reveal that DPP-PMRF significantly outperforms the OpenMP code, by amounts ranging from $2X$ to $7X$, depending upon the platform and concurrency level. 


The primary factor leading to this significant performance difference is the fact that the DPP formulation makes better use of the memory hierarchy. 
Whereas the OpenMP code operates in parallel over rows of a ragged array, DPP-PMRF recasts the problem as a series of atomic data parallel operations. 
To do so, it creates 1D arrays, which are then partitioned at runtime across a thread pool.
Such a formulation is much more amenable to vectorization, and results in significantly more uniform and predictable memory access patterns.
The size of these partitions, or chunks, is determined by TBB in a way to best match the cache characteristics and available parallelism of the underlying platform (see \S\ref{sec:results:software} for details). In contrast, the OpenMP code ``chunk size'' is the size of the given graph neighborhood being processed.
There is a significant performance difference that results when using a consistent and well chosen ``blocking factor,'' which results in better use of locality, in both space and time~\cite{Denning:2005}.
Our results are consistent with previous literature, which suggest one key factor to high performance on contemporary architectures is through code vectorization (c.f. Levesque and Vose, 2017~\cite{MPP-Programming:2017}). 
%

%
%

%
At higher levels of concurrency on the KNL platform, we see a significant performance decrease in DPP-PMRF, which is unexpected. At 64 cores, the runtime for DPP-PMRF actually \textit{increases} compared to the runtime for 32 cores.
Examination of detailed per-DPP timing indicates the runtime for two specific data parallel primitives, \textit{SortByKey} and \textit{ReduceByKey}, are the specific operations whose runtime increases going from 32 to 64 cores.
These data parallel primitives rely on an underlying vendor implementation in VTK-m with TBB as the back-end.
Further investigation is needed to better understand why these underlying vendor implementations decrease in performance going from 32 to 64 cores. 

\begin{figure*}[ht!]
\subfloat[Edison: Ivy Bridge]
{\includegraphics[width=0.45\linewidth]{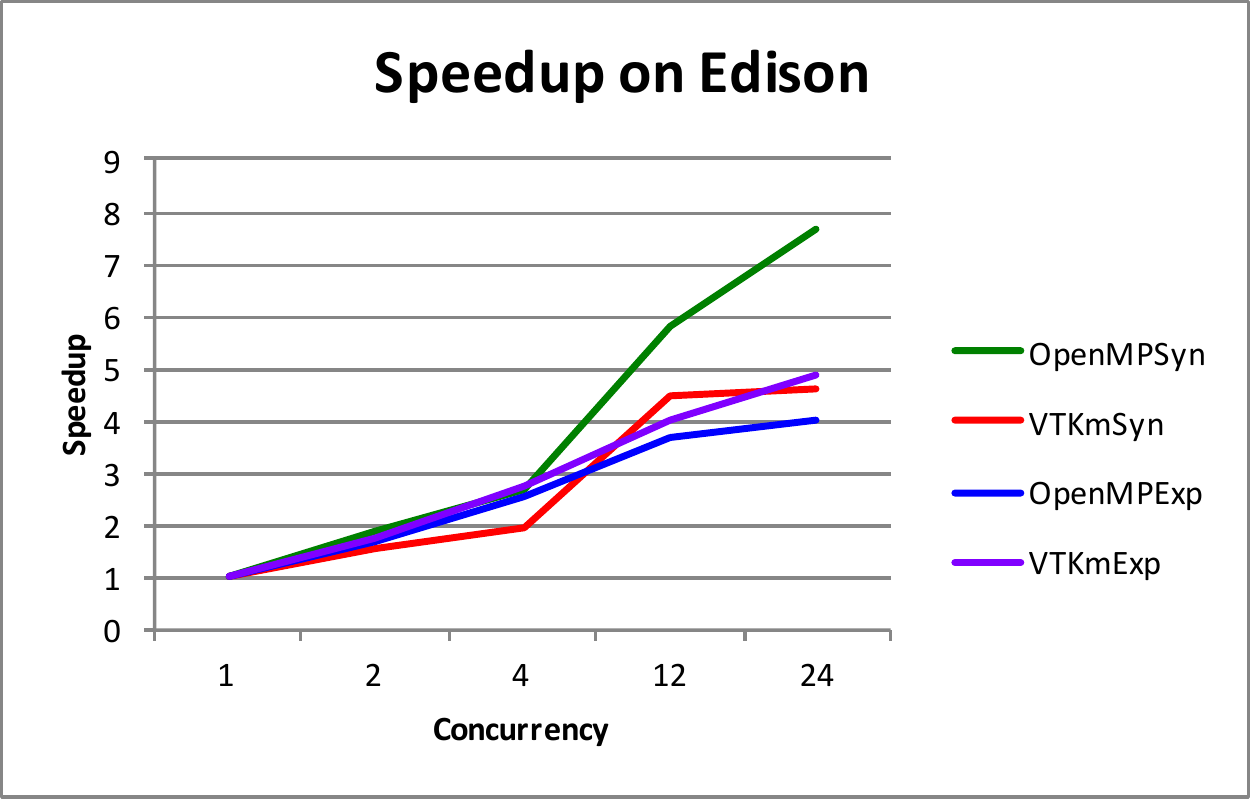}
\label{fig:edison-scaling}}
\hspace{2mm}
\subfloat[Cori: Knights Landing]
{\includegraphics[width=0.45\linewidth]{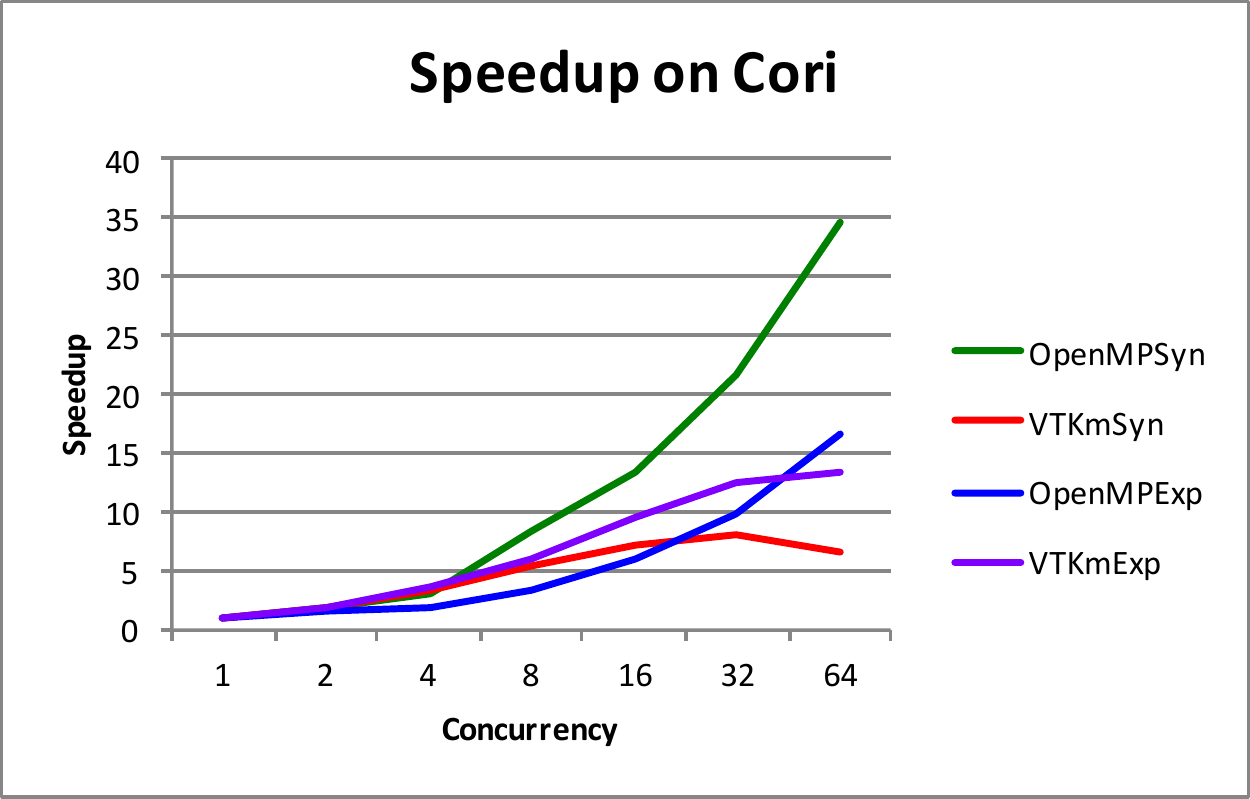}
\label{fig:cori-scaling}}
\caption{Speedup of the synthetic and experimental datasets on Edison and Cori. See \S\ref{sec:results:scalability} for more detail.}
\label{fig:ScalabilityCharts}
\end{figure*}

\subsubsection{Strong Scaling Results}
\label{sec:results:scalability}

The second performance study question we examine is the degree to which the OpenMP and DPP-PMRF implementations speed up with increasing concurrency. 
Holding the problem size fixed, we vary the concurrency on each platform, for each of the two image datasets.
Concurrency ranges from 1 to N, where N is the maximum number of cores on a node of each of the two platforms.

Results from this study are shown in Fig.~\ref{fig:ScalabilityCharts}, which show speedup curves for both implementations, on both platforms, at varying concurrency.
Looking at these results, the discussion that follows centers around two main themes.
First, how well are these codes scaling, and what might be the limits to scalability?
Second, how do scaling characteristics change with increasing concurrency, platform, and dataset?


In terms of how well these codes are scaling, the ideal rate of speedup would be equal to the number of cores: speedup of 2 on 2 cores, speedup of 4 on 4 cores, and so forth. 
The first observation is the both codes exhibit less than ideal scaling. The OpenMP code shows the best scaling on the synthetic dataset on both platforms, even though its absolute runtime is less than DPP-PMRF (except for one configuration, at 64 cores on the KNL platform).
The reasons for why these codes exhibit less than ideal scaling differ for each of the codes.

The OpenMP code, which uses loop-level parallelism over the neighborhoods of the graph, has as critical section that serializes access by all threads. 
This critical section is associated with a thread writing its results into an output buffer: each thread is updating a row of a ragged array.
We encountered what appears to be unexpected behavior with the C++ compiler on both platforms in which the output results were incorrect, unless this operation was serialized (see \S\ref{sec:results:software:openmp} for compilation details).
Future work will focus on eliminating this serial section of the code to improve scalability.

The DPP-PMRF code, which is a sequence of data parallel operations, depends upon an underlying vendor-provided implementation of key methods. 
In these studies, an analysis of runtime results looking at individual runtime for each of the DPP methods (detail not shown in Fig.~\ref{fig:ScalabilityCharts}), indicates that two specific DPP operations are limited in their scalability.
These two operations, a \textit{SortByKey} and \textit{ReduceByKey}, exhibit a maximum of about $5X$ speedup going from 1 to 24 cores on Edison, and about $11X$ speedup going from 1 to 64 cores on Cori.
As a result, the vendor-supplied implementation of the underlying DPP is in this case the limit to scalability.
We have observed in other studies looking at scalability of methods that use these same DPP on GPUs ~\cite{Lessley:LDAV17-1}, that the vendor-provided DPP implementation does not exhibit the same limit to scalability. In that study, the sort was being performed on arrays of integers. 
In the present study, we are sorting pairs of integers, which results in greater amounts of memory access and movement, more integer comparisons, as well as additional overhead to set up the arrays and work buffers for both those methods.


On Edison, both codes show a tail-off in speedup going from 12 to 24 cores.
%
%
A significant difference between these platforms is processor and memory speed: Edison has faster processors and slower memory; Cori has slower processors and faster memory.  The tail-off on Edison, for both codes, is most likely due to increasing memory pressure, as more cores are issuing an increasing number of memory operations. 


For the OpenMP code on both platforms, we see a noticeable difference in speedup curves for each of the two different datasets.
On both platforms, the OpenMP code scales better for the synthetic dataset.
Since the algorithm performance is a function of the complexity of the underlying data, specifically neighborhood size, we observe that these two datasets have vastly different demographics of neighborhood complexity (not shown due to space limitations).
In brief, the synthetic dataset has a larger number of smaller-sized neighborhoods and the histogram indicates bell-shaped distribution. In contrast, the experimental dataset has many more neighborhoods of higher complexity, and the distribution is very irregular. 
Because the OpenMP code parallelizes over individual neighborhoods, it is not possible to construct a workload distribution that attempts to create groups of neighborhoods that result in an even distribution of work across threads. 
We are relying on OpenMP's dynamic scheduling to achieve good load balance in the presence of an irregular workload distribution.
In this case, the result is the more irregular workload results in lower level of speedup for the OpenMP code on both platforms. 
In contrast, the DPP code reformulates this problem in a way that avoids this limitation.

\subsubsection{Platform Portability: GPU Results}

The final performance study question we examine
is an assessment of the platform-portable performance of DPP-PMRF.
We ran the algorithm on an NVIDIA Tesla K40 GPU accelerator, using the experimental and synthetic image datasets as input. 
The average GPU runtime for each dataset is compared to the average KNL CPU runtimes of both a serial (single core, hyper-threading disabled) execution of DPP-PMRF and the parallel execution of DPP-PMRF at maximum concurrency (68 cores, hyper-threading enabled; see \S\ref{sec:results:scalability}).

From~\autoref{table:gpuResults} we observe that, for both of the image datasets, DPP-PMRF achieves a significant speedup on the GPU over the serial version, with a maximum speedup of 44X on the experimental images.
Further, for both datasets, DPP-PMRF attains greater runtime performance on the GPU (maximum speedup of 13X on the experimental images), as compared to its execution on the KNL CPU platform.
These speedups demonstrate the ability of a GPU architecture to utilize the highly-parallel design of our algorithm, which consists of many fine-grained and compute-heavy data-parallel operations. 
Moreover, this experiment demonstrates the portable performance of DPP-PMRF, as we achieved improved runtimes without having to write custom, optimized NVIDIA CUDA GPU code within our algorithm; the same high-level algorithm was used for both the CPU and GPU experiments.

\begin{table}[t]
 \centering
 \caption{GPU runtimes (seconds) for DPP-PMRF over the experimental and synthetic image datasets, as compared to both serial and parallel CPU executions of DPP-PMRF on the KNL platform. The GPU speedup for a dataset is the serial CPU runtime divided by the DPP-PMRF GPU runtime. The CPU speedup is the DPP-PMRF CPU runtime divided by the DPP-PMRF GPU runtime.}
\begin{tabular}{ |p{2.5cm}||p{1.6cm}|p{1.3cm}| }
 \hline
 Platform / Dataset & Experimental & Synthetic\\
 \hline
 Serial CPU & 284.51 & 44.63 \\
 DPP-PMRF CPU & 22.77 & 7.09 \\
 DPP-PMRF GPU & 6.55 & 1.71 \\ 
 \hline \hline
 Speedup-CPU & 13X & 7X \\
 Speedup-GPU & 44X & 27X \\
 \hline
\end{tabular}
  \label{table:gpuResults}
\end{table}

\section{Conclusion and Future Work}

In the quest for platform portability and high performance, this work shows that reformulating 
a data-intensive graph problem using DPPs results in an implementation that achieves a significant speedup over a serial variant on both GPU and contemporary multi-core CPU architectures, and outperforms an OpenMP-parallel reference implementation.
%
The DPP-PMRF performance gain is the result of how it makes use of the memory hierarchy: it recasts the graph optimization into a series of 1D data-parallel operations, which are in turn more amenable for vectorization and fine-grained, thread-level parallelism on the underlying platform.
In contrast, the OpenMP implementation,
which is a coarse-grained, loop-parallelized version of  MRF optimization, has memory use patterns that are less efficient than the DPP-PMRF version.
%
Even though the OpenMP code is parallelized, ultimately its runtime performance is constrained by how it makes use of the memory hierarchy.

Our performance study focused on two multi-core platforms and a general-purpose GPU platform, and our results report total runtime. 
Future work will increase the diversity of platforms, to include newer GPU accelerators, as well as executing the OpenMP implementation on these GPU platforms for comparison with the DPP-PMRF GPU performance. The ability to dispatch OpenMP code to a GPU is an emerging capability, but not yet widespread (c.f.,~\cite{OpenMP:GPU:2018}).
Additionally, we will focus on collecting and analyzing additional hardware performance counters to gain a deeper understanding of memory utilization characteristics of both the OpenMP- and DPP-based codes. 

The method we present here operates on 2D images; the 3D datasets are processed as a stack of 2D images.
However, the PMRF method is, in theory, applicable to $n-$dimensional source data: the PMRF optimization takes a graph as input, and the dimensionality of the image isn't a factor once the MRF graph is constructed. 
We are currently extending our preprocessing pipeline to convert 3D structured images into an undirected graph format, which can enable DPP-PMRF to operate on 3D images directly, as opposed to a stack of 2D images. Moreover, we intend to extend the methods presented here to larger datasets, and to combine with a distributed-memory parallel PMRF~\cite{Heinemann:MPI-PMRF:2017} for a hybrid-parallel approach that is both platform-portable and that exhibits high performance for use on very large scientific data. 




%



\acknowledgments{
This work was supported by the Director, Office of Science, Office of Advanced Scientific Computing Research, of the U.S. Department of Energy under Contract No. DE-AC02-05CH11231, through the grants ``Towards Exascale: High Performance Visualization and Analytics,'' program manager Dr. Lucy Nowell, and "Scalable Data-Computing Convergence and Scientific Knowledge Discovery", program manager Dr. Laura Biven, and the Center for Applied Mathematics for Energy Related Applications (CAMERA).
This material is also based upon work supported by the U.S. Department of Energy, Office of Science, Office of Advanced
Scientific Computing Research, Award Number 14-017566, Program Manager Lucy Nowell.
This research used resources of the National Energy Research Scientific Computing Center, a DOE Office of Science User Facility supported by the Office of Science of the U.S. Department of Energy under Contract No. DE-AC02-05CH11231.
We also thank the LBNL-EESA division, especially P. Nico for the imaged geological samples.}

\bibliographystyle{abbrv-doi}

\bibliography{template}
\end{document}